\documentclass[prl,twocolumn,10pt,aps,longbibliography,nofootinbib,superscriptaddress]{revtex4-1}

\usepackage[utf8]{inputenc}
\usepackage{bm}
\usepackage{verbatim}
\usepackage{amsmath}
\usepackage{amssymb}
\usepackage{amsthm}
\usepackage{latexsym}
\usepackage{amsfonts}
\usepackage{epstopdf}
\usepackage{wasysym} % for certain special symbols
\usepackage{subfigure}
\usepackage{color}
\definecolor{darkblue}{rgb}{0,0,.5}
\definecolor{BLUE}{rgb}{0,0,1}
\definecolor{BLACK}{rgb}{0,0,0}
\usepackage[linktocpage, colorlinks=true, linkcolor=darkblue, citecolor=darkblue]{hyperref}
\usepackage[all]{hypcap}
\usepackage[makeroom]{cancel}
\usepackage{xfrac}
\usepackage{lipsum} % for leftskip command which allows to indent a paragraph

\newcommand{\C}[1]{{\cal{#1}}}

\newcommand{\bs}[1]{\boldsymbol{#1}}

\newcommand{\mf}[1]{{\mathfrak{#1}}}
\newcommand{\lr}[1]{{\langle {#1}\rangle}}

\newcommand{\rl}[0]{{\rangle\langle}}

\begin{document}
	
	\title{Toy Model Challenging Prevailing Definitions of Classicality}
	
	\author{Irene Valladares Duque}
	\affiliation{F\'isica Te\`orica: Informaci\'o i Fen\`omens Qu\`antics, Departament de F\'isica, Universitat Aut\`onoma de Barcelona, 08193 Bellaterra (Barcelona), Spain}
	\author{Philipp Strasberg}
	\affiliation{F\'isica Te\`orica: Informaci\'o i Fen\`omens Qu\`antics, Departament de F\'isica, Universitat Aut\`onoma de Barcelona, 08193 Bellaterra (Barcelona), Spain}
	\affiliation{Instituto de F\'isica de Cantabria (IFCA), Universidad de Cantabria--CSIC, 39005 Santander, Spain}

	\date{\today}
	
	\begin{abstract}
		We analyze a toy model that obeys environmentally induced decoherence and quantum Darwinism and satisfies the decoherent histories criterion and Leggett-Garg inequalities with respect to the pointer basis. Yet, the resulting ``classical'' dynamics are extremely fragile and recohere after a seemingly innocent control operation. This challenges the idea that classical behaviour or the Multiverse branches can be identified by looking at the universal wave function alone. It also demonstrates that quantum Darwinism is compatible with strong non-Markovianity. Possible solutions related to Markovianity, non-integrability and an operational definition of pointer states are briefly discussed.
	\end{abstract}
	
	\maketitle
	%\tableofcontents
	
	\newtheorem{mydef}{Definition}[section]
	\newtheorem*{mydef*}{Definition}
	\newtheorem{lemma}{Lemma}%[section]
	\newtheorem{conj}{Conjecture}
	\newtheorem{thm}{Theorem}[section]

\textbf{Introduction.}\,\,-- Several important theoretical concepts have been advanced to understand the emergence of classical behavior from microscopic quantum physics, such as environmentally induced decoherence (EID)~\cite{JoosEtAlBook2003, ZurekRMP2003, SchlosshauerPR2019}, quantum Darwinism (QD)~\cite{ZurekNP2009, KorbiczQuantum2021}, decoherent histories (DH)~\cite{GellMannHartleInBook1990, OmnesRMP1992, HalliwellANY1995, DowkerKentJSP1996} or Leggett-Garg inequalities (LGI)~\cite{EmaryLambertNoriRPP2014}. It is widespread to use words such as ``stable'', ``robust'', ``objective'', ``faithful'', ``(macro)realistic'', among others, to describe the resulting phenomenology. It is our objective here to show that these statements have to be taken \emph{cum grano salis} (with a grain of salt) and that not even the combination of EID, QD, DH and LGI is sufficient to explain ``stable, robust, etc.'' classical behavior.

Our results directly impact at least two research directions. First, there is growing interest in the many worlds interpretation of quantum mechanics~\cite{EverettRMP1957, DeWittPT1970, Vaidman2021}, which posits that the measurement postulate is superfluous and unitary evolution alone is sufficient to make sense of quantum mechanics. An essential question is then how the single, stable and robust world that we observe around us emerges from a coherent superposition of potentially very many different worlds. A crucial part of the answer is provided by the notions of classicality mentioned above, but as we show here these notions \emph{alone} can still imply a rather fragile classical world. In particular, recent research endeavours tried to infer classicality and/or the resulting branch structure of the Multiverse by looking \emph{solely} at the universal wave function~\cite{RiedelZurekZwolakPRA2016, RiedelPRL2017, WeingartenFP2022, OllivierEnt2022, TouilEtAlQuantum2024, TaylorMcCullochQuantum2025}. Our toy model highlights that this is likely too ambitious: details of the Hamiltonian can not be neglected to determine how stable, robust and classical a wave function branch is.
% Our results directly impact at least two research directions. First, there is growing interest in the many worlds interpretation of quantum mechanics~\cite{EverettRMP1957, DeWittPT1970, Vaidman2021}, and with it came the desire to read off classicality and/or the resulting branch structure of the Multiverse by looking \emph{solely} at the universal wave function~\cite{RiedelZurekZwolakPRA2016, RiedelPRL2017, WeingartenFP2022, OllivierEnt2022, TouilEtAlQuantum2024, TaylorMcCullochQuantum2025}.

Second, evidence has been collected that stimulated claims about quantum Darwinism implying Markovian behavior or non-Markovian behavior hindering quantum Darwinism~\cite{GiorgiGalveZambriniPRA2015, GalveZambriniManiscalco2016, PleasanceGarrawayPRA2017, MilazzoEtAlPRA2019, GuoHuangPLA2023}. We show these claims to be incorrect if one takes into account a rigorous definition of multi-time non-Markovianity~\cite{PollockEtAlPRL2018, LiHallWisemanPR2018, MilzModiPRXQ2021, accardi_quantum_1982}; for other counterarguments see also Ref.~\cite{LampoEtAlPRA2017, OliveiraPaulaDrumondPRA2019}.

In the remainder we analyze the toy model, demonstrate that it obeys EID, QD, DH and LGI, and show that the so identified classical degrees of freedom are fragile, sensitive and recoherent. Afterwards, we briefly discuss which additional ingredients could prevent such unwanted (because not observed in our real world) behavior. \\*

\textbf{Toy model.}\,\,-- We consider a spin system $\C S$ coupled to an environment $\C E$ composed of many particles $j\in\C E = \{1,\dots,N\}$ (we use $\C E$ both for the entire environment and as an index set). Their total Hamiltonian is taken to be
\begin{equation}\label{eq Hamiltonian}
 H = \frac{1}{2}\sigma_z\otimes\sum_{j\in \C E} g_j \hat q_j.
\end{equation}
Here, $\sigma_z$ is the Pauli matrix, $g_j\in\mathbb{R}$ the coupling strength and $\hat q_j$ the position operator of particle $j$. It is evident that this Hamiltonian is not realistic, but it is sufficient for a proof of principle that EID, QD, DH and LGI are not sufficient to describe a stable and realistic classical world. In particular, for a single environmental particle, $N=1$, this model has been analyzed before in the context of dynamical decoupling~\cite{PhysRevA.92.022102}, due to its counterintuitive non-Markovian properties~\cite{PollockEtAlPRL2018, SmirneEtAlQST2018, StrasbergDiazPRA2019, PhysRevX.10.041049} and it has been experimentally realized in Ref.~\cite{LiuEtAlNatPhys2011}. Moreover, we will discuss at the end that the here observed phenomenology emerges for a larger class of models.

The initial state is taken to be pure:
\begin{equation}
 |\Psi(0)\rangle = |\psi(0)\rangle_{\C S} \bigotimes_{j\in\C E} |\chi(0)\rangle_j.
\end{equation}
Here, $|\psi(0)\rangle_{\C S} = \alpha|0\rangle_{\C S} + \beta|1\rangle_{\C S}$ is the initial state of the system in the eigenbasis of $\sigma_z=|1\rl1|_{\C S}-|0\rl0|_{\C S}$, and the wave function of environmental particles in position representation is taken to be $\lr{q_j|\chi(0)}_j = \sqrt{\gamma_j/\pi}(q_j+i\gamma_j)^{-1}$, which implies a Lorentzian probability distribution with scale parameter $\gamma_j>0$. The exact time evolution of $|\Psi(t)\rangle = e^{-iHt}|\Psi(0)\rangle$ ($\hbar\equiv1$) is easily written down in position representation:
\begin{equation}\label{eq solution}
 \begin{split}
  |\Psi(t)\rangle =&~ \alpha|0\rangle_S \prod_{j\in\C E} \int dq_j \sqrt{\frac{\gamma_j}{\pi}}\frac{e^{ig_jq_jt/2}}{q_j+i\gamma_j} |q_j\rangle \\
  & + \beta|1\rangle_S \prod_{j\in\C E} \int dq_j \sqrt{\frac{\gamma_j}{\pi}}\frac{e^{-ig_jq_jt/2}}{q_j+i\gamma_j} |q_j\rangle.
 \end{split}
\end{equation}
Note that the $q_j$ appearing in the exponent is now a number and no longer an operator.

\begin{figure}[t]
 \centering\includegraphics[width=0.47\textwidth,clip=true]{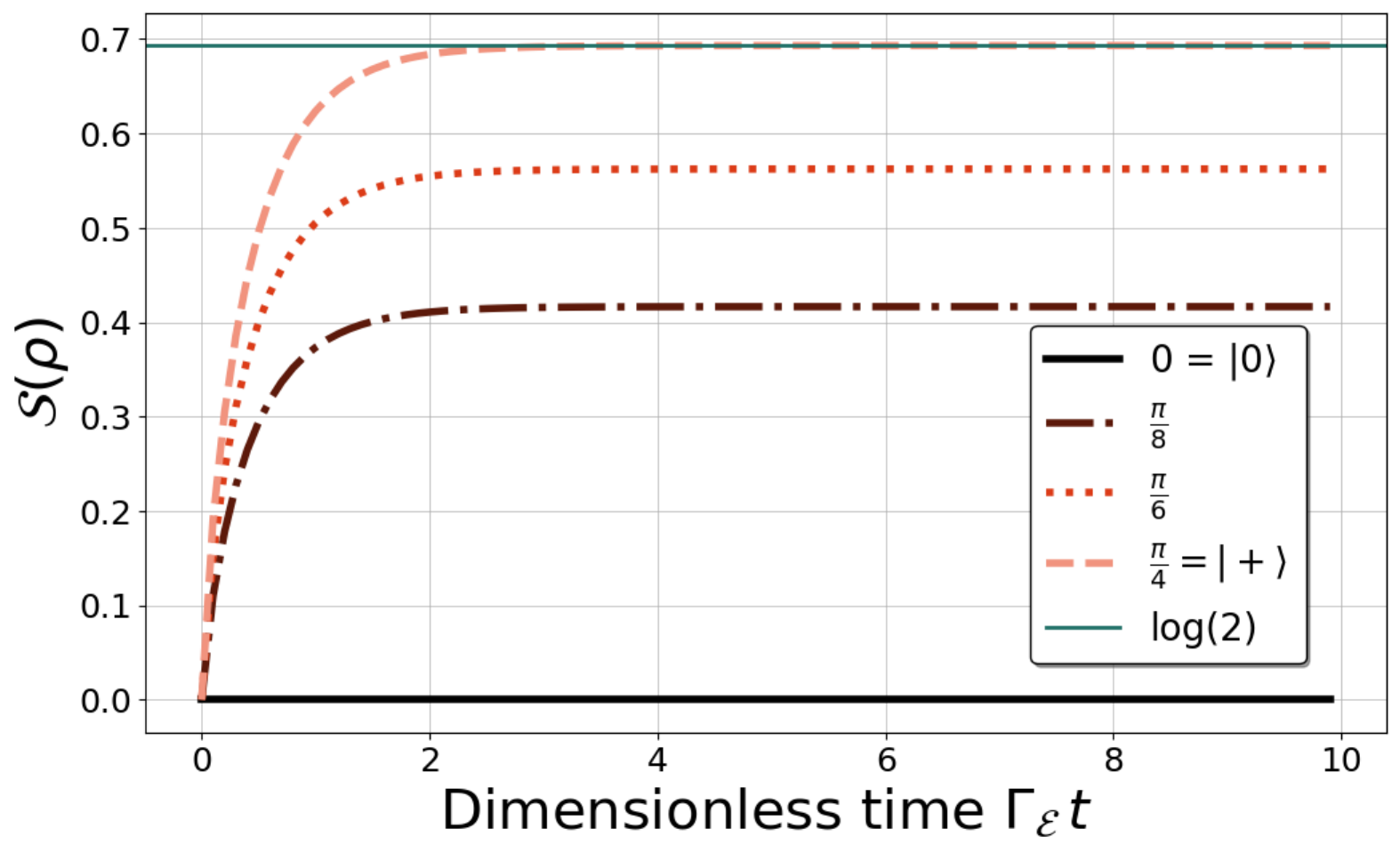}
 \label{fig 1}
 \caption{Numerical test of the predictability sieve by plotting the time evolution of von Neumann entropy for different initial states $|\psi(0)\rangle = e^{-i\phi\sigma_y}|0\rangle$ with $\phi\in\{0,\pi/8,\pi/6,\pi/4\}$. We see a rapid rise of entropy, i.e., loss of predictability, for all states except the pointer state $|0\rangle$. Parameters of the model are $g_j = 1$ and $\gamma_j = 1$ for all $j$. }
\end{figure}

We confirm that the model obeys EID and has a well defined pointer basis. To this end, consider the quantum master equation for the open system,
\begin{equation}\label{eq QME}
 \frac{\partial}{\partial t}\rho_S(t) = \C L\rho_S(t) \equiv \frac{\Gamma_{\C E}}{2} \big[\sigma_z\rho_S(t)\sigma_z - \rho_S(t)\big],
\end{equation}
where $\Gamma_{\C E} = \sum_{j\in\C E} g_j\gamma_j$ is a rate. The \emph{exact} validity of eqn~(\ref{eq QME}) can be confirmed by directly comparing its solution $\rho_{\C S}(t) = e^{\C Lt}\rho_{\C S}(0)$ with $\mbox{tr}_{\C E}\{|\Psi(t)\rl\Psi(t)|\}$ for $\rho_{\C S}(0) = |\psi(0)\rl\psi(0)|_{\C S}$. Now, recall that the pointer states in an open quantum system (if they exist) identify the states in which \emph{any} initial density matrix quickly looses its off-diagonal terms (or ``coherences''), which, by looking at eqn~(\ref{eq QME}), is here evidently the eigenbasis of $\sigma_z$: $|0\rangle_{\C S}$ and $|1\rangle_{\C S}$. The pointer states can be also identified with Zurek's predictability sieve~\cite{ZurekPTP1993} by checking which pure states remain predictable---in the sense of suffering the least change in von Neumann entropy---under time evolution, as depicted in Fig.~\ref{fig 1}.
% It is evident that the eigenstates of $\sigma_z$, $|0\rangle_{\C S}$ and $|1\rangle_{\C S}$, are the pointer states of the system. This fact can be also confirmed using Zurek's predictability sieve~\cite{ZurekPTP1993}, see Fig.~\ref{fig 1}.

We continue by confirming QD. To this end, we split the environment $\C E = \C F \otimes \bar{\C F}$ into two fragments $\C F = \{1,\dots,L\}$ and $\bar{\C F} = \{L+1,\dots,N\}$. Assuming that neither $\C F$ nor $\bar{\C F}$ are very small, QD relies on the idea that the overall (universal) wave function can---at least to a good approximation---be written as
\begin{equation}\label{eq branching form}
 |\Psi\rangle = \sum_z \sqrt{p_z}|\psi_z\rangle_{\C S}|\phi_z\rangle_{\C F}|\varphi_z\rangle_{\bar{\C F}},
\end{equation}
where $\{p_z\}$ is a set of probabilities, $\{|\psi_z\rangle_{\C S}\}$ is the orthonormal pointer basis of the system, and $\{|\phi_z\rangle_{\C F}\}$ $(\{|\varphi_z\rangle_{\bar{\C F}}\})$ is an orthonormal set of vectors in $\C F$ $(\bar{\C F})$. Equation~(\ref{eq branching form}) is also known as the \emph{branching form} of the wave function: in essence, QD posits that the Schmidt decomposition---which always exists for any bipartition---remains valid for all tripartitions $\C S\otimes\C F\otimes\bar{\C F}$ with $\C F, \bar{\C F}$ not unreasonably small.\footnote{Strictly speaking, Eq.~(\ref{eq branching form}) implies a so-called spectrum broadcast structure, which is a stronger notion than the initial idea of QD~\cite{HorodeckiKorbiczHorodeckiPRA2015, LeOlayaCastroPRL2019, KorbiczQuantum2021}.}. 

Comparing eqns~(\ref{eq solution}) and~(\ref{eq branching form}) suggests the identification $z\in\{0,1\}$, $|\psi_0\rangle_{\C S} = |0\rangle_{\C S}$, $|\psi_1\rangle_{\C S} = |1\rangle_{\C S}$ and
\begin{align}
 |\phi_0(t)\rangle_{\C F} &= \prod_{j\in\C F} \int dq_j \sqrt{\frac{\gamma_j}{\pi}}\frac{e^{ig_jq_jt/2}}{q_j+i\gamma_j} |q_j\rangle, \\
 |\phi_1(t)\rangle_{\C F} &= \prod_{j\in\C F} \int dq_j \sqrt{\frac{\gamma_j}{\pi}}\frac{e^{-ig_jq_jt/2}}{q_j+i\gamma_j} |q_j\rangle.
\end{align}
Note that both expressions only differ by the sign of the exponent and that an obvious analogous definition applies to $\bar{\C F}$. It is easy to confirm normalization, $\lr{\phi_z(t)|\phi_z(t)}_{\C F} = 1$, and their overlap decays with rate $\Gamma_{\C F} = \sum_{j\in\C F} g_j\gamma_j$:
\begin{equation}
 \lr{\phi_0(t)|\phi_1(t)}_{\C F} = e^{-\Gamma_{\C F}t}.
\end{equation}
Thus, convergence to the branching form happens exponentially fast for all possible (non-trivial) fragments, implying QD.

Next, we confirm DH in the pointer basis. To this end, we denote by $\Pi_z = |z\rl z|_{\C S}\otimes 1_{\C E}$ ($z\in\{0,1\}$) projectors on the pointer basis and introduce the unitary time evolution operator $U_{\ell,k} = e^{-iH(t_\ell-t_k)}$ from time $t_k$ to $t_\ell$. For histories $\bs z = (z_n,\dots,z_1)$ we then introduce history states
\begin{equation}
 |\psi(\bs z)\rangle = \Pi_{z_n}U_{n,n-1}\cdots \Pi_{z_1}U_{1,0}|\psi(0)\rangle,
\end{equation}
which condition the wave function on passing through subspaces $z_i$ at time $t_i$. The DH condition reads
\begin{equation}
 \lr{\psi(\bs z)|\psi(\bs z')} = 0 ~~~ \text{for all} ~~~ \bs z\neq\bs z',
\end{equation}
which is easily confirmed by noting that $[\Pi_z,U_{\ell,k}] = 0$. Finally, the validity of LGI follows from the validity of the DH condition.

\begin{figure}[!t]
 \centering\includegraphics[width=0.47\textwidth,clip=true]{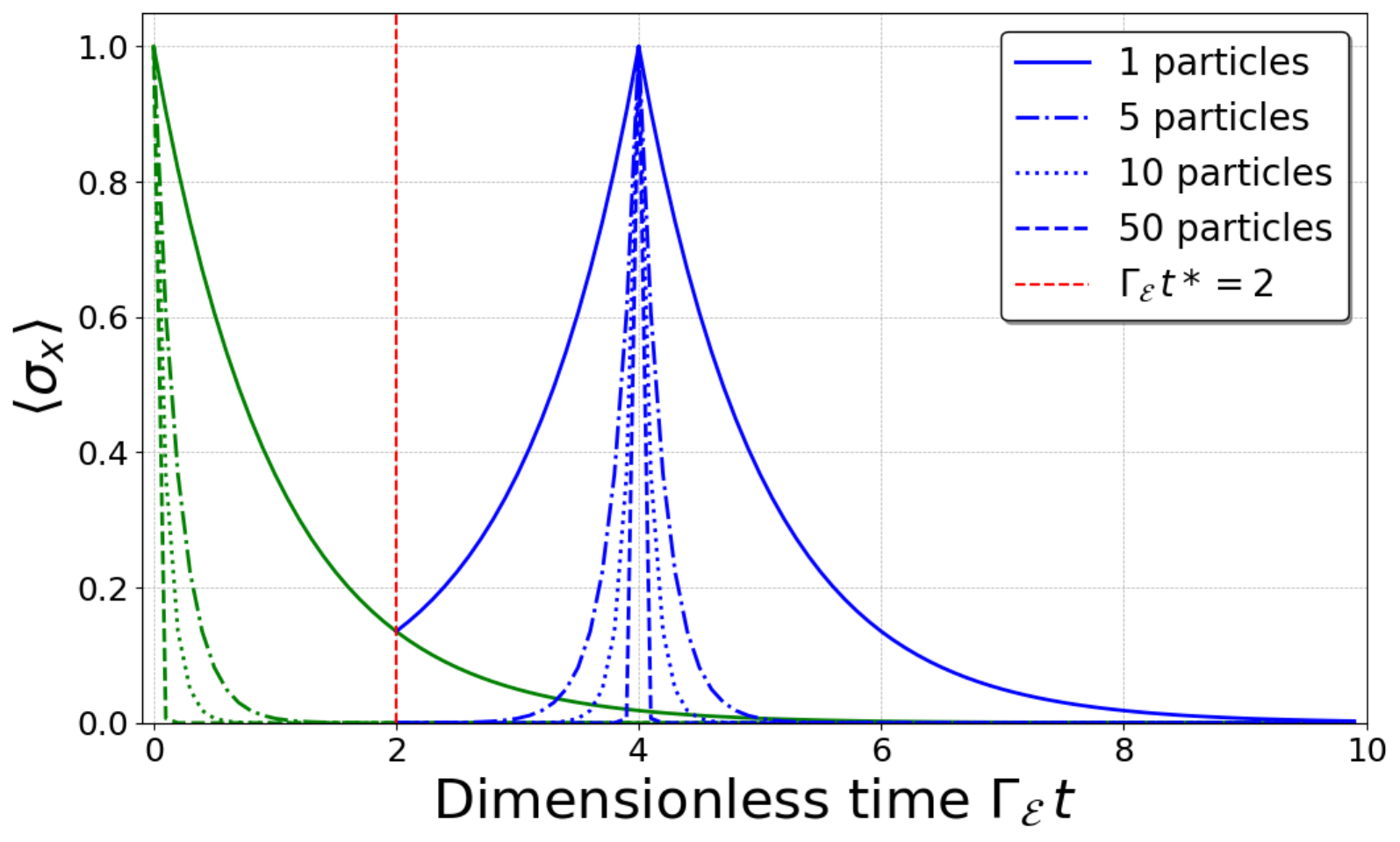}
 \label{fig 2}
 \caption{Effect of the control operation $\mf C$ on the coherences $\lr{\sigma_x}(t)$ as a function of time $t$ for different numbers of environmental particles. $\mf C$ happens at $\Gamma_{\C E}t^*=2$ (dashed vertical red line) and causes perfect recoherence at $t = 2t^*$ (blue lines). The green lines display the evolution without $\mf C$. The initial state was $|\psi(0)\rangle_{\C S} = |+\rangle_{\C S} \equiv (|0\rangle_{\C S}+|1\rangle_{\C S})/\sqrt{2}$ and model parameters are as in Fig.~\ref{fig 1}. }
\end{figure}

Thus, the system satisfies EID, QD, DH and LGI, but the resulting dynamics of the so-identified classical degrees of freedom is extremely fragile. To see this, we apply a spin flip control operation $\mf C$ to $\C S$, defined by its action on the open system as $\mf C\rho_{\C S}(t^*) \equiv \sigma_x\rho_{\C S}(t^*)\sigma_x$ with $\sigma_x = |0\rl 1|_{\C S} + |1\rl0|_{\C S}$, at an arbitrary time $t^*$. As exemplified in Fig.~\ref{fig 2}, the effect of this control operation is that the system evolves back to its initial state. 
In particular, any coherences in the pointer basis, which were lost during the interval $[0,t^*]$ and distributed among the environmental particles in unison with QD, recohere during the interval $[t^*,2t^*]$: \emph{classical information becomes quantum information again}.

It is remarkable that such drastic changes in the dynamics are caused by an innocent control operation $\mf C$. It acts only locally, i.e., we do not assume any unrealistic control over the entire quantum system. Moreover, for the maximally coherent state $|\psi(0)\rangle_{\C S} = |+\rangle_{\C S}$, it does not even change the system state, $\mf C\rho_{\C S}(t) = \rho_{\C S}(t)$, but still causes a complete recoherence. In addition, it is not necessary that an outside experimenter performs $\mf C$: within a purely unitarily evolving quantum Universe, $\mf C$ could be caused by another particle flying past system $\C S$. Finally, any other control operation $\mf C'\rho_{\C S}(t) = \sum_\alpha K_\alpha\rho_{\C S}(t) K_\alpha^\dagger$, which contains Kraus operators $K_\alpha$ related to $\sigma_x$, will cause some partial recoherence. Thus, this phenomenology can certainly not be called stable, robust, faithful, objective, etc., as the usual meaning of these words suggests that small perturbations can not cause such drastic changes.

Looking for a cause of this drastic effect, we find that
\begin{equation}\label{eq time reversal}
 \sigma_x e^{-iHt}\sigma_x = e^{+iHt}
\end{equation}
and conclude that $\mf C$ effectively implements a \emph{time reversal} of the $\C S\C E$-dynamics---very similar to a spin (Hahn) echo effect~\cite{HahnPR1950}: $|\Psi(0)\rangle = \sigma_x U_{t,0}\sigma_x U_{t,0} |\Psi(0)\rangle$. After realizing this, it becomes clear that we can induce recoherences as often as we like and at arbitrary times. This also does not depend on the specific initial wavefunction of the environmental particles, which we chose here only for convenience. Moreover, we can further conclude that the here uncovered fragile nature of classical information is not tied to the specific Hamiltonian~(\ref{eq Hamiltonian}). We could replace the environmental particles by any other sort of quantum system (e.g., spins) and would find the same phenomenology. We could also include a self Hamiltonian $H_{\C S}$ and $H_{\C E}$ in the description as long as it commutes with the interaction Hamiltonian.

Finally, it is interesting to note that the notion of pointer states and QD are sensitive to this unstable behaviour in the sense that Zurek's predictability sieve temporarily breaks down and QD temporarily ceases to exist. In contrast, if one considers histories of the form
\begin{equation}
 |\psi(\bs z)\rangle = \Pi_{z_n}U_{n,n-1}\cdots\sigma_x\cdots \Pi_{z_1}U_{1,0}|\psi(0)\rangle
\end{equation}
with the $\sigma_x$ rotation happening at an arbitrary point in time, one can show with the help of eqn~(\ref{eq time reversal}) that the criteria for DH (and thus also LGI) remain \emph{intact}.\footnote{When considering $\lr{\psi(\bs z)|\psi(\bs z')}$ one can commute both $\sigma_x$ through until they meet, giving rise to an identity $\sigma_x^2 = I$. In between, this flips the projectors (i.e., $\sigma_x\Pi_{0}\sigma_x = \Pi_{1}$) and inverts the unitary time evolution operators, which, however, continue to commute with the projectors, and the DH condition follows.} \\*

\textbf{Discussion.}\,\,-- In our daily life such unstable and fragile classical behavior is not observed. But if EID, QD, DH and LGI can not prevent this sort of behavior, what else could prevent it? We briefly discuss three possible answers.

One answer to the question invokes Markovianity. Our toy model is non-Markovian due to a (rather extreme) breakdown of the quantum regression theorem~\cite{PollockEtAlPRL2018, LiHallWisemanPR2018, MilzModiPRXQ2021}: while a Markovian master equation exactly describes the evolution of $\rho_S(t)$, eqn~(\ref{eq QME}) can not be used to compute correlation functions for system operators $A, B$:
\begin{equation}
 \begin{split}
  \lr{A(t)B(s)} &\equiv \mbox{tr}\{A U_{t,s} B U_{s,0}|\Psi(0)\rl\Psi(0)|U^\dagger_{s,0}U^\dagger_{t,s}\} \\
  &\neq \mbox{tr}\{Ae^{\C L(t-s)}Be^{\C Ls}\rho_S(0)\}.
 \end{split}
\end{equation}
% Thus, does the existence of a classical world requires Markovianity? Probably not, given that there exist also classical non-Markovian processes. Moreover, invoking Markovianity seems unsatisfactory from a foundational point of view as long as microscopically proving multi-time Markovianity remains a challenging task~\cite{VanKampenPhys1954, DuemckeJMP1983, FordOConnellPRL1996, FigueroaRomeroModiPollockQuantum2019, FigueroaRomeroPollockModiCP2021, StrasbergEtAlPRA2023}.
Thus, does the existence of a classical world require Markovianity? Probably not, given that there exist also classical non-Markovian processes. Moreover, invoking Markovianity seems unsatisfactory from a foundational point of view since deriving microscopically multi-time Markovianity from first principles remains a challenging task~\cite{VanKampenPhys1954, DuemckeJMP1983, FordOConnellPRL1996, FigueroaRomeroModiPollockQuantum2019, FigueroaRomeroPollockModiCP2021, StrasbergEtAlPRA2023}.

Another answer, which seems more feasible and realistic, invokes non-integrability. In fact, that a control operation acting only on the open system has such a drastic effect on the entire dynamics is likely an artifact of the simplicity of the model, which is integrable: it has an extensive number of conserved quantities ($\sigma_z$ and all the $\hat q_j$) and the eigenstates and eigenvalues of $H$ can be immediately written down. Any generic random and small perturbation added to the model will destroy these symmetries and make it very likely non-integrable---while the question as to how strong this perturbation needs to be to completey suppress the here observed phenomenology remains an interesting research direction. In any case, the robustness of non-integrable models to local perturbations and their fast scrambling of information among many degrees of freedom makes them certainly good candidates for a truly robust emergence of classical behavior, as recently highlighted within the context of DH by one of the authors~\cite{StrasbergEtAlPRA2023, StrasbergSP2023, StrasbergReinhardSchindlerPRX2024, WangStrasbergArXiv2024} (though the necessity of non-integrability has been conjectured by Van Kampen~\cite{VanKampenPhys1954}).

Finally, it remains to answer how one could experimentally decide whether the pointer states of a given open quantum system are robust and stable: as we saw above, Zurek's predictability sieve does not answer this question, and neither does QD, DH or LGI. To this end, we like to propose the following operational definition:

\begin{mydef*}[$(\epsilon,\tau)$-pointer states]
 Consider an open quantum system ${\C S}$ and an arbitrary sequence of control operations $(\mf C_n,\dots,\mf C_1)$, restricted only to act locally on ${\C S}$, at arbitrary but fixed times $\{t_n,\dots,t_1\}$. An orthogonal basis $\{|\phi_n\rangle_{\C S}\}$ (or, more generally, a complete orthogonal set of projectors $\{\Pi_n\}$) is called $(\epsilon,\tau)$-pointer states if
 \begin{equation}
  \frac{1}{2}\|\rho_{\C S}(t;\mf C_n,\dots,\mf C_1) - \Delta\rho_{\C S}(t;\mf C_n,\dots,\mf C_1)\|_1 \le \epsilon
 \end{equation}
 for all $n\ge1$ and all times $t \ge t_n+\tau$. Here, $\|\cdot\|_1$ is the trace norm, $\rho_{\C S}(t;\mf C_n,\dots,\mf C_1)$ the normalized state conditional on applying $(\mf C_n,\dots,\mf C_1)$, and $\Delta\rho_{\C S}\equiv \sum_n\Pi_n\rho_{\C S}\Pi_n$ the dephasing operation with respect to the above basis.
 \end{mydef*}

The above definition encodes the idea that---on an experimental resolution scale $\epsilon$ and a temporal resolution scale $\tau$ (note the similarity to $(\epsilon,\tau)$-entropy~\cite{GaspardWangPR1993})---the existence of a good pointer basis causes the actual state of the open quantum system to be indistinguishable from a dephased state, indepedent of which local control operation one has applied to the open system. Studying the detailed consequences of this definition remains an intriguing task for the future, but we can note already here that the above toy model does not allow to identify a reasonable $(\epsilon,\tau)$-pointer basis.\\*

\textbf{Conclusion.}\,\,-- Our model raises awareness of the fact that EID, QD, DH and LGI are not sufficient to explain the emergence of a classical world, and in particular they alone do not justify widespread assertions about stability, robustness, objectivity, etc. We emphasize that our findings shall \emph{not} imply that EID, QD, DH and LGI are useless concepts: they are certainly among the most important ideas we have to talk about the emergence of classicality. Instead, our results should serve as a reminder that nature is subtle, and complex topics such as the emergence of classicality or the branches of the Multiverse can not be understood from a single and narrow perspective. To counterbalance this tendency, we proposed additional mechanisms complementing EID, QD, DH and LGI. In particular, non-integrability and operational approaches appear to us as promising candidates to get a better fundamental and practical understanding of the problem.

In addition to its implications for current research, we find that the here studied toy model is of great pedagogical value to illustrate counterintuitive properties of open quantum systems in relation to master equations, (non-)Markovianity, various notions of classicality and operational approaches to quantum dynamics. For this reason the model might deserve to be more widely known.

Finally, whether the here theoretically calculated phenomenology can be demonstrated in real experiments, such as modern spin echo implementations~\cite{DebnathEtAlPRL2020, WeichselbaumerEtAlPRL2020}, remains a fascinating question.

\acknowledgments
Discussions with Joey Schindler are gratefully acknowledged. Financial support by MICINN with funding from European Union NextGenerationEU (PRTR-C17.I1) and by the Generalitat de Catalunya (project 2017-SGR-1127) are acknowledged. PS is further supported by ``la Caixa'' Foundation (ID 100010434, fellowship code LCF/BQ/PR21/11840014), the Ram\'on y Cajal program RYC2022-035908-I, the European Commission QuantERA grant ExTRaQT (Spanish MICIN project PCI2022-132965), and the Spanish MINECO (project PID2019-107609GB-I00) with the support of FEDER funds.\\*

%\emph{Data availability statement:} No new data were created or analysed in this study.
%\bibliographystyle{eplbib}
\bibliography{classicality_v4.bib}

\begin{thebibliography}{51}%
\makeatletter
\providecommand \@ifxundefined [1]{%
 \@ifx{#1\undefined}
}%
\providecommand \@ifnum [1]{%
 \ifnum #1\expandafter \@firstoftwo
 \else \expandafter \@secondoftwo
 \fi
}%
\providecommand \@ifx [1]{%
 \ifx #1\expandafter \@firstoftwo
 \else \expandafter \@secondoftwo
 \fi
}%
\providecommand \natexlab [1]{#1}%
\providecommand \enquote  [1]{``#1''}%
\providecommand \bibnamefont  [1]{#1}%
\providecommand \bibfnamefont [1]{#1}%
\providecommand \citenamefont [1]{#1}%
\providecommand \href@noop [0]{\@secondoftwo}%
\providecommand \href [0]{\begingroup \@sanitize@url \@href}%
\providecommand \@href[1]{\@@startlink{#1}\@@href}%
\providecommand \@@href[1]{\endgroup#1\@@endlink}%
\providecommand \@sanitize@url [0]{\catcode `\\12\catcode `\$12\catcode
  `\&12\catcode `\#12\catcode `\^12\catcode `\_12\catcode `\%12\relax}%
\providecommand \@@startlink[1]{}%
\providecommand \@@endlink[0]{}%
\providecommand \url  [0]{\begingroup\@sanitize@url \@url }%
\providecommand \@url [1]{\endgroup\@href {#1}{\urlprefix }}%
\providecommand \urlprefix  [0]{URL }%
\providecommand \Eprint [0]{\href }%
\providecommand \doibase [0]{http://dx.doi.org/}%
\providecommand \selectlanguage [0]{\@gobble}%
\providecommand \bibinfo  [0]{\@secondoftwo}%
\providecommand \bibfield  [0]{\@secondoftwo}%
\providecommand \translation [1]{[#1]}%
\providecommand \BibitemOpen [0]{}%
\providecommand \bibitemStop [0]{}%
\providecommand \bibitemNoStop [0]{.\EOS\space}%
\providecommand \EOS [0]{\spacefactor3000\relax}%
\providecommand \BibitemShut  [1]{\csname bibitem#1\endcsname}%
\let\auto@bib@innerbib\@empty
%</preamble>
\bibitem [{\citenamefont {Joos}\ \emph {et~al.}(2003)\citenamefont {Joos},
  \citenamefont {Zeh}, \citenamefont {Kiefer}, \citenamefont {Giulini},
  \citenamefont {Kupsch},\ and\ \citenamefont {Stamatescu}}]{JoosEtAlBook2003}%
  \BibitemOpen
  \bibfield  {author} {\bibinfo {author} {\bibfnamefont {E.}~\bibnamefont
  {Joos}}, \bibinfo {author} {\bibfnamefont {H.~D.}\ \bibnamefont {Zeh}},
  \bibinfo {author} {\bibfnamefont {C.}~\bibnamefont {Kiefer}}, \bibinfo
  {author} {\bibfnamefont {D.}~\bibnamefont {Giulini}}, \bibinfo {author}
  {\bibfnamefont {J.}~\bibnamefont {Kupsch}}, \ and\ \bibinfo {author}
  {\bibfnamefont {I.-O.}\ \bibnamefont {Stamatescu}},\ }\href@noop {} {\emph
  {\bibinfo {title} {Decoherence and the Appearance of a Classical World in
  Quantum Theory}}}\ (\bibinfo  {publisher} {Springer},\ \bibinfo {address}
  {Berlin Heidelberg},\ \bibinfo {year} {2003})\BibitemShut {NoStop}%
\bibitem [{\citenamefont {Zurek}(2003)}]{ZurekRMP2003}%
  \BibitemOpen
  \bibfield  {author} {\bibinfo {author} {\bibfnamefont {W.~H.}\ \bibnamefont
  {Zurek}},\ }\bibfield  {title} {\enquote {\bibinfo {title} {Decoherence,
  einselection, and the quantum origins of the classical},}\ }\href {\doibase
  10.1103/RevModPhys.75.715} {\bibfield  {journal} {\bibinfo  {journal}
  {Reviews of Modern Physics}\ }\textbf {\bibinfo {volume} {75}},\ \bibinfo
  {pages} {715--775} (\bibinfo {year} {2003})}\BibitemShut {NoStop}%
\bibitem [{\citenamefont {Schlosshauer}(2019)}]{SchlosshauerPR2019}%
  \BibitemOpen
  \bibfield  {author} {\bibinfo {author} {\bibfnamefont {M.}~\bibnamefont
  {Schlosshauer}},\ }\bibfield  {title} {\enquote {\bibinfo {title} {Quantum
  decoherence},}\ }\href {\doibase 10.1016/j.physrep.2019.10.001} {\bibfield
  {journal} {\bibinfo  {journal} {Physics Reports}\ }\textbf {\bibinfo {volume}
  {831}},\ \bibinfo {pages} {1--57} (\bibinfo {year} {2019})}\BibitemShut
  {NoStop}%
\bibitem [{\citenamefont {Zurek}(2009)}]{ZurekNP2009}%
  \BibitemOpen
  \bibfield  {author} {\bibinfo {author} {\bibfnamefont {W.~H.}\ \bibnamefont
  {Zurek}},\ }\bibfield  {title} {\enquote {\bibinfo {title} {Quantum
  darwinism},}\ }\href {\doibase 10.1038/nphys1202} {\bibfield  {journal}
  {\bibinfo  {journal} {Nature Physics}\ }\textbf {\bibinfo {volume} {5}}
  (\bibinfo {year} {2009}),\ 10.1038/nphys1202}\BibitemShut {NoStop}%
\bibitem [{\citenamefont {Korbicz}(2021)}]{KorbiczQuantum2021}%
  \BibitemOpen
  \bibfield  {author} {\bibinfo {author} {\bibfnamefont {J.~K.}\ \bibnamefont
  {Korbicz}},\ }\bibfield  {title} {\enquote {\bibinfo {title} {Roads to
  objectivity: Quantum darwinism, spectrum broadcast structures, and strong
  quantum darwinism – a review},}\ }\href {\doibase
  10.22331/q-2021-11-08-571} {\bibfield  {journal} {\bibinfo  {journal}
  {Quantum}\ }\textbf {\bibinfo {volume} {5}},\ \bibinfo {pages} {571}
  (\bibinfo {year} {2021})}\BibitemShut {NoStop}%
\bibitem [{\citenamefont {Gell-Mann}\ and\ \citenamefont
  {Hartle}(1990)}]{GellMannHartleInBook1990}%
  \BibitemOpen
  \bibfield  {author} {\bibinfo {author} {\bibfnamefont {M.}~\bibnamefont
  {Gell-Mann}}\ and\ \bibinfo {author} {\bibfnamefont {J.~B.}\ \bibnamefont
  {Hartle}},\ }\bibfield  {title} {\enquote {\bibinfo {title} {Quantum
  mechanics in the light of quantum cosmology},}\ }in\ \href@noop {} {\emph
  {\bibinfo {booktitle} {Complexity, Entropy and the Physics of Information}}}\
  (\bibinfo  {publisher} {Addison-Wesley},\ \bibinfo {address} {Reading},\
  \bibinfo {year} {1990})\ pp.\ \bibinfo {pages} {425--459}\BibitemShut
  {NoStop}%
\bibitem [{\citenamefont {Omn\`es}(1992)}]{OmnesRMP1992}%
  \BibitemOpen
  \bibfield  {author} {\bibinfo {author} {\bibfnamefont {R.}~\bibnamefont
  {Omn\`es}},\ }\bibfield  {title} {\enquote {\bibinfo {title} {Consistent
  interpretations of quantum mechanics},}\ }\href {\doibase
  10.1103/RevModPhys.64.339} {\bibfield  {journal} {\bibinfo  {journal}
  {Reviews of Modern Physics}\ }\textbf {\bibinfo {volume} {64}},\ \bibinfo
  {pages} {339--382} (\bibinfo {year} {1992})}\BibitemShut {NoStop}%
\bibitem [{\citenamefont {Halliwell}(1995)}]{HalliwellANY1995}%
  \BibitemOpen
  \bibfield  {author} {\bibinfo {author} {\bibfnamefont {J.~J.}\ \bibnamefont
  {Halliwell}},\ }\bibfield  {title} {\enquote {\bibinfo {title} {A review of
  the decoherent histories approach to quantum mechanics},}\ }\href {\doibase
  10.1111/j.1749-6632.1995.tb39014.x} {\bibfield  {journal} {\bibinfo
  {journal} {Annals of the New York Academy of Sciences}\ }\textbf {\bibinfo
  {volume} {755}},\ \bibinfo {pages} {726--740} (\bibinfo {year}
  {1995})}\BibitemShut {NoStop}%
\bibitem [{\citenamefont {Dowker}\ and\ \citenamefont
  {Kent}(1996)}]{DowkerKentJSP1996}%
  \BibitemOpen
  \bibfield  {author} {\bibinfo {author} {\bibfnamefont {F.}~\bibnamefont
  {Dowker}}\ and\ \bibinfo {author} {\bibfnamefont {A.}~\bibnamefont {Kent}},\
  }\bibfield  {title} {\enquote {\bibinfo {title} {On the consistent histories
  approach to quantum mechanics},}\ }\href {\doibase 10.1007/BF02183396}
  {\bibfield  {journal} {\bibinfo  {journal} {Journal of Statistical Physics}\
  }\textbf {\bibinfo {volume} {82}} (\bibinfo {year} {1996}),\
  10.1007/BF02183396}\BibitemShut {NoStop}%
\bibitem [{\citenamefont {Emary}\ \emph {et~al.}(2014)\citenamefont {Emary},
  \citenamefont {Lambert},\ and\ \citenamefont
  {Nori}}]{EmaryLambertNoriRPP2014}%
  \BibitemOpen
  \bibfield  {author} {\bibinfo {author} {\bibfnamefont {C.}~\bibnamefont
  {Emary}}, \bibinfo {author} {\bibfnamefont {N.}~\bibnamefont {Lambert}}, \
  and\ \bibinfo {author} {\bibfnamefont {F.}~\bibnamefont {Nori}},\ }\bibfield
  {title} {\enquote {\bibinfo {title} {Leggett-garg inequalities},}\ }\href
  {\doibase 10.1088/0034-4885/77/1/016001} {\bibfield  {journal} {\bibinfo
  {journal} {Reports on Progress in Physics}\ }\textbf {\bibinfo {volume}
  {77}},\ \bibinfo {pages} {039501} (\bibinfo {year} {2014})}\BibitemShut
  {NoStop}%
\bibitem [{\citenamefont {Everett}(1957)}]{EverettRMP1957}%
  \BibitemOpen
  \bibfield  {author} {\bibinfo {author} {\bibfnamefont {H.}~\bibnamefont
  {Everett}},\ }\bibfield  {title} {\enquote {\bibinfo {title} {Relative state
  formulation of quantum mechanics},}\ }\href {\doibase
  10.1103/RevModPhys.29.454} {\bibfield  {journal} {\bibinfo  {journal}
  {Reviews of Modern Physics}\ }\textbf {\bibinfo {volume} {29}},\ \bibinfo
  {pages} {454--462} (\bibinfo {year} {1957})}\BibitemShut {NoStop}%
\bibitem [{\citenamefont {Witt}(1970)}]{DeWittPT1970}%
  \BibitemOpen
  \bibfield  {author} {\bibinfo {author} {\bibfnamefont {B.~S.~De}\
  \bibnamefont {Witt}},\ }\bibfield  {title} {\enquote {\bibinfo {title}
  {Quantum mechanics and reality},}\ }\href {\doibase 10.1063/1.3022331}
  {\bibfield  {journal} {\bibinfo  {journal} {Physics Today}\ }\textbf
  {\bibinfo {volume} {23}},\ \bibinfo {pages} {30} (\bibinfo {year}
  {1970})}\BibitemShut {NoStop}%
\bibitem [{\citenamefont {Vaidman}(2021)}]{Vaidman2021}%
  \BibitemOpen
  \bibfield  {author} {\bibinfo {author} {\bibfnamefont {L.}~\bibnamefont
  {Vaidman}},\ }\bibfield  {title} {\enquote {\bibinfo {title} {Many-worlds
  interpretation of quantum mechanics},}\ }in\ \href@noop {} {\emph {\bibinfo
  {booktitle} {Stanford Encyclopedia of Philosophy}}}\ (\bibinfo {year}
  {2021})\ \bibinfo {edition} {fall 2021}\ ed.,\ \bibinfo {note}
  {https://plato.stanford.edu/archives/fall2021/entries/qm-manyworlds/}\BibitemShut
  {NoStop}%
\bibitem [{\citenamefont {Riedel}\ \emph {et~al.}(2016)\citenamefont {Riedel},
  \citenamefont {Zurek},\ and\ \citenamefont
  {Zwolak}}]{RiedelZurekZwolakPRA2016}%
  \BibitemOpen
  \bibfield  {author} {\bibinfo {author} {\bibfnamefont {C.~J.}\ \bibnamefont
  {Riedel}}, \bibinfo {author} {\bibfnamefont {W.~H.}\ \bibnamefont {Zurek}}, \
  and\ \bibinfo {author} {\bibfnamefont {M.}~\bibnamefont {Zwolak}},\
  }\bibfield  {title} {\enquote {\bibinfo {title} {Objective past of a quantum
  universe: Redundant records of consistent histories},}\ }\href {\doibase
  10.1103/PhysRevA.93.032126} {\bibfield  {journal} {\bibinfo  {journal}
  {Physical Review A}\ }\textbf {\bibinfo {volume} {93}},\ \bibinfo {pages}
  {032126} (\bibinfo {year} {2016})}\BibitemShut {NoStop}%
\bibitem [{\citenamefont {Riedel}(2017)}]{RiedelPRL2017}%
  \BibitemOpen
  \bibfield  {author} {\bibinfo {author} {\bibfnamefont {C.~Jess}\ \bibnamefont
  {Riedel}},\ }\bibfield  {title} {\enquote {\bibinfo {title} {Classical branch
  structure from spatial redundancy in a many-body wave function},}\ }\href
  {\doibase 10.1103/PhysRevLett.118.120402} {\bibfield  {journal} {\bibinfo
  {journal} {Physical Review Letters}\ }\textbf {\bibinfo {volume} {118}},\
  \bibinfo {pages} {120402} (\bibinfo {year} {2017})}\BibitemShut {NoStop}%
\bibitem [{\citenamefont {Weingarten}(2022)}]{WeingartenFP2022}%
  \BibitemOpen
  \bibfield  {author} {\bibinfo {author} {\bibfnamefont {D.}~\bibnamefont
  {Weingarten}},\ }\bibfield  {title} {\enquote {\bibinfo {title} {Macroscopic
  reality from quantum complexity},}\ }\href {\doibase
  10.1007/s10701-022-00554-0} {\bibfield  {journal} {\bibinfo  {journal}
  {Foundations of Physics}\ }\textbf {\bibinfo {volume} {52}},\ \bibinfo
  {pages} {45} (\bibinfo {year} {2022})}\BibitemShut {NoStop}%
\bibitem [{\citenamefont {Ollivier}(2022)}]{OllivierEnt2022}%
  \BibitemOpen
  \bibfield  {author} {\bibinfo {author} {\bibfnamefont {H.}~\bibnamefont
  {Ollivier}},\ }\bibfield  {title} {\enquote {\bibinfo {title} {Emergence of
  objectivity for quantum many-body systems},}\ }\href {\doibase
  10.3390/e24020277} {\bibfield  {journal} {\bibinfo  {journal} {Entropy}\
  }\textbf {\bibinfo {volume} {24}},\ \bibinfo {pages} {277} (\bibinfo {year}
  {2022})}\BibitemShut {NoStop}%
\bibitem [{\citenamefont {Touil}\ \emph {et~al.}(2024)\citenamefont {Touil},
  \citenamefont {Anza}, \citenamefont {Deffner},\ and\ \citenamefont
  {Crutchfield}}]{TouilEtAlQuantum2024}%
  \BibitemOpen
  \bibfield  {author} {\bibinfo {author} {\bibfnamefont {A.}~\bibnamefont
  {Touil}}, \bibinfo {author} {\bibfnamefont {F.}~\bibnamefont {Anza}},
  \bibinfo {author} {\bibfnamefont {S.}~\bibnamefont {Deffner}}, \ and\
  \bibinfo {author} {\bibfnamefont {J.~P.}\ \bibnamefont {Crutchfield}},\
  }\bibfield  {title} {\enquote {\bibinfo {title} {Branching states as the
  emergent structure of a quantum universe},}\ }\href {\doibase
  10.22331/q-2024-10-10-1494} {\bibfield  {journal} {\bibinfo  {journal}
  {Quantum}\ }\textbf {\bibinfo {volume} {8}},\ \bibinfo {pages} {1494}
  (\bibinfo {year} {2024})}\BibitemShut {NoStop}%
\bibitem [{\citenamefont {Taylor}\ and\ \citenamefont
  {McCulloch}(2025)}]{TaylorMcCullochQuantum2025}%
  \BibitemOpen
  \bibfield  {author} {\bibinfo {author} {\bibfnamefont {J.~K.}\ \bibnamefont
  {Taylor}}\ and\ \bibinfo {author} {\bibfnamefont {I.~P.}\ \bibnamefont
  {McCulloch}},\ }\bibfield  {title} {\enquote {\bibinfo {title} {Wavefunction
  branching: when you can't tell pure states from mixed states},}\ }\href
  {\doibase 10.22331/q-2025-03-25-1670} {\bibfield  {journal} {\bibinfo
  {journal} {Quantum}\ }\textbf {\bibinfo {volume} {9}},\ \bibinfo {pages}
  {1670} (\bibinfo {year} {2025})}\BibitemShut {NoStop}%
\bibitem [{\citenamefont {Giorgi}\ \emph {et~al.}(2015)\citenamefont {Giorgi},
  \citenamefont {Galve},\ and\ \citenamefont
  {Zambrini}}]{GiorgiGalveZambriniPRA2015}%
  \BibitemOpen
  \bibfield  {author} {\bibinfo {author} {\bibfnamefont {G.~L.}\ \bibnamefont
  {Giorgi}}, \bibinfo {author} {\bibfnamefont {F.}~\bibnamefont {Galve}}, \
  and\ \bibinfo {author} {\bibfnamefont {R.}~\bibnamefont {Zambrini}},\
  }\bibfield  {title} {\enquote {\bibinfo {title} {Quantum darwinism and
  non-markovian dissipative dynamics from quantum phases of the spin-1/2 xx
  model},}\ }\href {\doibase 10.1103/PhysRevA.92.022105} {\bibfield  {journal}
  {\bibinfo  {journal} {Phys. Rev. A}\ }\textbf {\bibinfo {volume} {92}},\
  \bibinfo {pages} {022105} (\bibinfo {year} {2015})}\BibitemShut {NoStop}%
\bibitem [{\citenamefont {Galve}\ \emph {et~al.}(2016)\citenamefont {Galve},
  \citenamefont {Zambrini},\ and\ \citenamefont
  {Maniscalco}}]{GalveZambriniManiscalco2016}%
  \BibitemOpen
  \bibfield  {author} {\bibinfo {author} {\bibfnamefont {F.}~\bibnamefont
  {Galve}}, \bibinfo {author} {\bibfnamefont {R.}~\bibnamefont {Zambrini}}, \
  and\ \bibinfo {author} {\bibfnamefont {S.}~\bibnamefont {Maniscalco}},\
  }\bibfield  {title} {\enquote {\bibinfo {title} {Non-markovianity hinders
  quantum darwinism},}\ }\href {\doibase 10.1038/srep19607} {\bibfield
  {journal} {\bibinfo  {journal} {Sci. Rep.}\ }\textbf {\bibinfo {volume}
  {6}},\ \bibinfo {pages} {19607} (\bibinfo {year} {2016})}\BibitemShut
  {NoStop}%
\bibitem [{\citenamefont {Pleasance}\ and\ \citenamefont
  {Garraway}(2017)}]{PleasanceGarrawayPRA2017}%
  \BibitemOpen
  \bibfield  {author} {\bibinfo {author} {\bibfnamefont {G.}~\bibnamefont
  {Pleasance}}\ and\ \bibinfo {author} {\bibfnamefont {B.~M.}\ \bibnamefont
  {Garraway}},\ }\bibfield  {title} {\enquote {\bibinfo {title} {Application of
  quantum darwinism to a structured environment},}\ }\href {\doibase
  10.1103/PhysRevA.96.062105} {\bibfield  {journal} {\bibinfo  {journal} {Phys.
  Rev. A}\ }\textbf {\bibinfo {volume} {96}},\ \bibinfo {pages} {062105}
  (\bibinfo {year} {2017})}\BibitemShut {NoStop}%
\bibitem [{\citenamefont {Milazzo}\ \emph {et~al.}(2019)\citenamefont
  {Milazzo}, \citenamefont {Lorenzo}, \citenamefont {Paternostro},\ and\
  \citenamefont {Palma}}]{MilazzoEtAlPRA2019}%
  \BibitemOpen
  \bibfield  {author} {\bibinfo {author} {\bibfnamefont {N.}~\bibnamefont
  {Milazzo}}, \bibinfo {author} {\bibfnamefont {S.}~\bibnamefont {Lorenzo}},
  \bibinfo {author} {\bibfnamefont {M.}~\bibnamefont {Paternostro}}, \ and\
  \bibinfo {author} {\bibfnamefont {G.~Massimo}\ \bibnamefont {Palma}},\
  }\bibfield  {title} {\enquote {\bibinfo {title} {Role of information backflow
  in the emergence of quantum darwinism},}\ }\href {\doibase
  10.1103/PhysRevA.100.012101} {\bibfield  {journal} {\bibinfo  {journal}
  {Phys. Rev. A}\ }\textbf {\bibinfo {volume} {100}},\ \bibinfo {pages}
  {012101} (\bibinfo {year} {2019})}\BibitemShut {NoStop}%
\bibitem [{\citenamefont {Guo}\ and\ \citenamefont
  {Huang}(2023)}]{GuoHuangPLA2023}%
  \BibitemOpen
  \bibfield  {author} {\bibinfo {author} {\bibfnamefont {X.-K.}\ \bibnamefont
  {Guo}}\ and\ \bibinfo {author} {\bibfnamefont {Z.}~\bibnamefont {Huang}},\
  }\bibfield  {title} {\enquote {\bibinfo {title} {On the relation between
  quantum darwinism and approximate quantum markovianity},}\ }\href {\doibase
  10.1016/j.physleta.2023.129204} {\bibfield  {journal} {\bibinfo  {journal}
  {Phys. Lett. A}\ }\textbf {\bibinfo {volume} {491}},\ \bibinfo {pages}
  {129204} (\bibinfo {year} {2023})}\BibitemShut {NoStop}%
\bibitem [{\citenamefont {Pollock}\ \emph {et~al.}(2018)\citenamefont
  {Pollock}, \citenamefont {Rodr{\'i}guez-Rosario}, \citenamefont {Frauenheim},
  \citenamefont {Paternostro},\ and\ \citenamefont
  {Modi}}]{PollockEtAlPRL2018}%
  \BibitemOpen
  \bibfield  {author} {\bibinfo {author} {\bibfnamefont {F.~A.}\ \bibnamefont
  {Pollock}}, \bibinfo {author} {\bibfnamefont {C.}~\bibnamefont
  {Rodr{\'i}guez-Rosario}}, \bibinfo {author} {\bibfnamefont {T.}~\bibnamefont
  {Frauenheim}}, \bibinfo {author} {\bibfnamefont {M.}~\bibnamefont
  {Paternostro}}, \ and\ \bibinfo {author} {\bibfnamefont {K.}~\bibnamefont
  {Modi}},\ }\bibfield  {title} {\enquote {\bibinfo {title} {Operational markov
  condition for quantum processes},}\ }\href {\doibase
  10.1103/PhysRevLett.120.040405} {\bibfield  {journal} {\bibinfo  {journal}
  {Phys. Rev. Lett.}\ }\textbf {\bibinfo {volume} {120}},\ \bibinfo {pages}
  {040405} (\bibinfo {year} {2018})}\BibitemShut {NoStop}%
\bibitem [{\citenamefont {Li}\ \emph {et~al.}(2018)\citenamefont {Li},
  \citenamefont {Hall},\ and\ \citenamefont {Wiseman}}]{LiHallWisemanPR2018}%
  \BibitemOpen
  \bibfield  {author} {\bibinfo {author} {\bibfnamefont {L.}~\bibnamefont
  {Li}}, \bibinfo {author} {\bibfnamefont {M.~J.~W.}\ \bibnamefont {Hall}}, \
  and\ \bibinfo {author} {\bibfnamefont {H.~M.}\ \bibnamefont {Wiseman}},\
  }\bibfield  {title} {\enquote {\bibinfo {title} {Concepts of quantum
  non-markovianity: A hierarchy},}\ }\href {\doibase
  10.1016/j.physrep.2018.07.001} {\bibfield  {journal} {\bibinfo  {journal}
  {Phys. Rep.}\ }\textbf {\bibinfo {volume} {759}},\ \bibinfo {pages} {1--51}
  (\bibinfo {year} {2018})}\BibitemShut {NoStop}%
\bibitem [{\citenamefont {Milz}\ and\ \citenamefont
  {Modi}(2021)}]{MilzModiPRXQ2021}%
  \BibitemOpen
  \bibfield  {author} {\bibinfo {author} {\bibfnamefont {S.}~\bibnamefont
  {Milz}}\ and\ \bibinfo {author} {\bibfnamefont {K.}~\bibnamefont {Modi}},\
  }\bibfield  {title} {\enquote {\bibinfo {title} {Quantum stochastic processes
  and quantum non-markovian phenomena},}\ }\href {\doibase
  10.1103/PRXQuantum.2.030201} {\bibfield  {journal} {\bibinfo  {journal} {PRX
  Quantum}\ }\textbf {\bibinfo {volume} {2}},\ \bibinfo {pages} {030201}
  (\bibinfo {year} {2021})}\BibitemShut {NoStop}%
\bibitem [{\citenamefont {Accardi}\ \emph {et~al.}(1982)\citenamefont
  {Accardi}, \citenamefont {Frigerio},\ and\ \citenamefont
  {Lewis}}]{accardi_quantum_1982}%
  \BibitemOpen
  \bibfield  {author} {\bibinfo {author} {\bibfnamefont {Luigi}\ \bibnamefont
  {Accardi}}, \bibinfo {author} {\bibfnamefont {Alberto}\ \bibnamefont
  {Frigerio}}, \ and\ \bibinfo {author} {\bibfnamefont {John~T.}\ \bibnamefont
  {Lewis}},\ }\bibfield  {title} {\enquote {\bibinfo {title} {Quantum
  {Stochastic} {Processes}},}\ }\href {\doibase 10.2977/prims/1195184017}
  {\bibfield  {journal} {\bibinfo  {journal} {Publ. Res. Inst. Math. Sci.}\
  }\textbf {\bibinfo {volume} {18}},\ \bibinfo {pages} {97--133} (\bibinfo
  {year} {1982})}\BibitemShut {NoStop}%
\bibitem [{\citenamefont {Lampo}\ \emph {et~al.}(2017)\citenamefont {Lampo},
  \citenamefont {Tuziemski}, \citenamefont {Lewenstein},\ and\ \citenamefont
  {Korbicz}}]{LampoEtAlPRA2017}%
  \BibitemOpen
  \bibfield  {author} {\bibinfo {author} {\bibfnamefont {A.}~\bibnamefont
  {Lampo}}, \bibinfo {author} {\bibfnamefont {J.}~\bibnamefont {Tuziemski}},
  \bibinfo {author} {\bibfnamefont {M.}~\bibnamefont {Lewenstein}}, \ and\
  \bibinfo {author} {\bibfnamefont {J.~K.}\ \bibnamefont {Korbicz}},\
  }\bibfield  {title} {\enquote {\bibinfo {title} {Objectivity in the
  non-markovian spin-boson model},}\ }\href {\doibase
  10.1103/PhysRevA.96.012120} {\bibfield  {journal} {\bibinfo  {journal} {Phys.
  Rev. A}\ }\textbf {\bibinfo {volume} {96}},\ \bibinfo {pages} {012120}
  (\bibinfo {year} {2017})}\BibitemShut {NoStop}%
\bibitem [{\citenamefont {Oliveira}\ \emph {et~al.}(2019)\citenamefont
  {Oliveira}, \citenamefont {de~Paula},\ and\ \citenamefont
  {Drumond}}]{OliveiraPaulaDrumondPRA2019}%
  \BibitemOpen
  \bibfield  {author} {\bibinfo {author} {\bibfnamefont {S.~M.}\ \bibnamefont
  {Oliveira}}, \bibinfo {author} {\bibfnamefont {A.~L.}\ \bibnamefont
  {de~Paula}}, \ and\ \bibinfo {author} {\bibfnamefont {R.~C.}\ \bibnamefont
  {Drumond}},\ }\bibfield  {title} {\enquote {\bibinfo {title} {Quantum
  darwinism and non-markovianity in a model of quantum harmonic oscillators},}\
  }\href {\doibase 10.1103/PhysRevA.100.052110} {\bibfield  {journal} {\bibinfo
   {journal} {Phys. Rev. A}\ }\textbf {\bibinfo {volume} {100}},\ \bibinfo
  {pages} {052110} (\bibinfo {year} {2019})}\BibitemShut {NoStop}%
\bibitem [{\citenamefont {Arenz}\ \emph {et~al.}(2015)\citenamefont {Arenz},
  \citenamefont {Hillier}, \citenamefont {Fraas},\ and\ \citenamefont
  {Burgarth}}]{PhysRevA.92.022102}%
  \BibitemOpen
  \bibfield  {author} {\bibinfo {author} {\bibfnamefont {Christian}\
  \bibnamefont {Arenz}}, \bibinfo {author} {\bibfnamefont {Robin}\ \bibnamefont
  {Hillier}}, \bibinfo {author} {\bibfnamefont {Martin}\ \bibnamefont {Fraas}},
  \ and\ \bibinfo {author} {\bibfnamefont {Daniel}\ \bibnamefont {Burgarth}},\
  }\bibfield  {title} {\enquote {\bibinfo {title} {Distinguishing decoherence
  from alternative quantum theories by dynamical decoupling},}\ }\href
  {\doibase 10.1103/PhysRevA.92.022102} {\bibfield  {journal} {\bibinfo
  {journal} {Phys. Rev. A}\ }\textbf {\bibinfo {volume} {92}},\ \bibinfo
  {pages} {022102} (\bibinfo {year} {2015})}\BibitemShut {NoStop}%
\bibitem [{\citenamefont {Smirne}\ \emph {et~al.}(2018)\citenamefont {Smirne},
  \citenamefont {Egloff}, \citenamefont {Díaz}, \citenamefont {Plenio},\ and\
  \citenamefont {Hulega}}]{SmirneEtAlQST2018}%
  \BibitemOpen
  \bibfield  {author} {\bibinfo {author} {\bibfnamefont {A.}~\bibnamefont
  {Smirne}}, \bibinfo {author} {\bibfnamefont {D.}~\bibnamefont {Egloff}},
  \bibinfo {author} {\bibfnamefont {M.~G.}\ \bibnamefont {Díaz}}, \bibinfo
  {author} {\bibfnamefont {M.~B.}\ \bibnamefont {Plenio}}, \ and\ \bibinfo
  {author} {\bibfnamefont {S.~F.}\ \bibnamefont {Hulega}},\ }\bibfield  {title}
  {\enquote {\bibinfo {title} {Coherence and non-classicality of quantum markov
  processes},}\ }\href {\doibase 10.1088/2058-9565/aaebd5} {\bibfield
  {journal} {\bibinfo  {journal} {Quantum Sci. Technol.}\ }\textbf {\bibinfo
  {volume} {4}},\ \bibinfo {pages} {01LT01} (\bibinfo {year}
  {2018})}\BibitemShut {NoStop}%
\bibitem [{\citenamefont {Strasberg}\ and\ \citenamefont
  {Díaz}(2019)}]{StrasbergDiazPRA2019}%
  \BibitemOpen
  \bibfield  {author} {\bibinfo {author} {\bibfnamefont {P.}~\bibnamefont
  {Strasberg}}\ and\ \bibinfo {author} {\bibfnamefont {M.~G.}\ \bibnamefont
  {Díaz}},\ }\bibfield  {title} {\enquote {\bibinfo {title} {Classical quantum
  stochastic processes},}\ }\href {\doibase 10.1103/PhysRevA.100.022120}
  {\bibfield  {journal} {\bibinfo  {journal} {Phys. Rev. A}\ }\textbf {\bibinfo
  {volume} {100}},\ \bibinfo {pages} {022120} (\bibinfo {year}
  {2019})}\BibitemShut {NoStop}%
\bibitem [{\citenamefont {Milz}\ \emph {et~al.}(2020)\citenamefont {Milz},
  \citenamefont {Egloff}, \citenamefont {Taranto}, \citenamefont {Theurer},
  \citenamefont {Plenio}, \citenamefont {Smirne},\ and\ \citenamefont
  {Huelga}}]{PhysRevX.10.041049}%
  \BibitemOpen
  \bibfield  {author} {\bibinfo {author} {\bibfnamefont {Simon}\ \bibnamefont
  {Milz}}, \bibinfo {author} {\bibfnamefont {Dario}\ \bibnamefont {Egloff}},
  \bibinfo {author} {\bibfnamefont {Philip}\ \bibnamefont {Taranto}}, \bibinfo
  {author} {\bibfnamefont {Thomas}\ \bibnamefont {Theurer}}, \bibinfo {author}
  {\bibfnamefont {Martin~B.}\ \bibnamefont {Plenio}}, \bibinfo {author}
  {\bibfnamefont {Andrea}\ \bibnamefont {Smirne}}, \ and\ \bibinfo {author}
  {\bibfnamefont {Susana~F.}\ \bibnamefont {Huelga}},\ }\bibfield  {title}
  {\enquote {\bibinfo {title} {When is a non-markovian quantum process
  classical?}}\ }\href {\doibase 10.1103/PhysRevX.10.041049} {\bibfield
  {journal} {\bibinfo  {journal} {Phys. Rev. X}\ }\textbf {\bibinfo {volume}
  {10}},\ \bibinfo {pages} {041049} (\bibinfo {year} {2020})}\BibitemShut
  {NoStop}%
\bibitem [{\citenamefont {Liu}\ \emph {et~al.}(2011)\citenamefont {Liu},
  \citenamefont {Li}, \citenamefont {Huang}, \citenamefont {Li}, \citenamefont
  {Guo}, \citenamefont {Laine}, \citenamefont {Breuer},\ and\ \citenamefont
  {Piilo}}]{LiuEtAlNatPhys2011}%
  \BibitemOpen
  \bibfield  {author} {\bibinfo {author} {\bibfnamefont {B.-H.}\ \bibnamefont
  {Liu}}, \bibinfo {author} {\bibfnamefont {L.}~\bibnamefont {Li}}, \bibinfo
  {author} {\bibfnamefont {Y.-F.}\ \bibnamefont {Huang}}, \bibinfo {author}
  {\bibfnamefont {C.-F.}\ \bibnamefont {Li}}, \bibinfo {author} {\bibfnamefont
  {G.-C.}\ \bibnamefont {Guo}}, \bibinfo {author} {\bibfnamefont {E.-M.}\
  \bibnamefont {Laine}}, \bibinfo {author} {\bibfnamefont {H.-P.}\ \bibnamefont
  {Breuer}}, \ and\ \bibinfo {author} {\bibfnamefont {J.}~\bibnamefont
  {Piilo}},\ }\bibfield  {title} {\enquote {\bibinfo {title} {Experimental
  control of the transition from markovian to non-markovian dynamics of open
  quantum systems},}\ }\href@noop {} {\bibfield  {journal} {\bibinfo  {journal}
  {Nat. Phys.}\ }\textbf {\bibinfo {volume} {7}},\ \bibinfo {pages} {931--934}
  (\bibinfo {year} {2011})}\BibitemShut {NoStop}%
\bibitem [{\citenamefont {Zurek}(1993)}]{ZurekPTP1993}%
  \BibitemOpen
  \bibfield  {author} {\bibinfo {author} {\bibfnamefont {W.~H.}\ \bibnamefont
  {Zurek}},\ }\bibfield  {title} {\enquote {\bibinfo {title} {Preferred states,
  predictability, classicality and the environment-induced decoherence},}\
  }\href {\doibase 10.1143/ptp/89.2.281} {\bibfield  {journal} {\bibinfo
  {journal} {Prog. Theor. Phys.}\ }\textbf {\bibinfo {volume} {89}},\ \bibinfo
  {pages} {281--312} (\bibinfo {year} {1993})}\BibitemShut {NoStop}%
\bibitem [{\citenamefont {Horodecki}\ \emph {et~al.}(2015)\citenamefont
  {Horodecki}, \citenamefont {Korbicz},\ and\ \citenamefont
  {Horodecki}}]{HorodeckiKorbiczHorodeckiPRA2015}%
  \BibitemOpen
  \bibfield  {author} {\bibinfo {author} {\bibfnamefont {R.}~\bibnamefont
  {Horodecki}}, \bibinfo {author} {\bibfnamefont {J.~K.}\ \bibnamefont
  {Korbicz}}, \ and\ \bibinfo {author} {\bibfnamefont {P.}~\bibnamefont
  {Horodecki}},\ }\bibfield  {title} {\enquote {\bibinfo {title} {Quantum
  origins of objectivity},}\ }\href {\doibase 10.1103/PhysRevA.91.032122}
  {\bibfield  {journal} {\bibinfo  {journal} {Phys. Rev. A}\ }\textbf {\bibinfo
  {volume} {91}},\ \bibinfo {pages} {032122} (\bibinfo {year}
  {2015})}\BibitemShut {NoStop}%
\bibitem [{\citenamefont {Le}\ and\ \citenamefont
  {Olaya-Castro}(2019)}]{LeOlayaCastroPRL2019}%
  \BibitemOpen
  \bibfield  {author} {\bibinfo {author} {\bibfnamefont {T.~P.}\ \bibnamefont
  {Le}}\ and\ \bibinfo {author} {\bibfnamefont {A.}~\bibnamefont
  {Olaya-Castro}},\ }\bibfield  {title} {\enquote {\bibinfo {title} {Strong
  quantum darwinism and strong independence are equivalent to spectrum
  broadcast structure},}\ }\href {\doibase 10.1103/PhysRevLett.122.010403}
  {\bibfield  {journal} {\bibinfo  {journal} {Phys. Rev. Lett.}\ }\textbf
  {\bibinfo {volume} {122}},\ \bibinfo {pages} {010403} (\bibinfo {year}
  {2019})}\BibitemShut {NoStop}%
\bibitem [{\citenamefont {Hahn}(1950)}]{HahnPR1950}%
  \BibitemOpen
  \bibfield  {author} {\bibinfo {author} {\bibfnamefont {E.~L.}\ \bibnamefont
  {Hahn}},\ }\bibfield  {title} {\enquote {\bibinfo {title} {Spin echoes},}\
  }\href {\doibase 10.1103/PhysRev.80.580} {\bibfield  {journal} {\bibinfo
  {journal} {Phys. Rev.}\ }\textbf {\bibinfo {volume} {80}},\ \bibinfo {pages}
  {580--594} (\bibinfo {year} {1950})}\BibitemShut {NoStop}%
\bibitem [{\citenamefont {Kampen}(1954)}]{VanKampenPhys1954}%
  \BibitemOpen
  \bibfield  {author} {\bibinfo {author} {\bibfnamefont {N.~Van}\ \bibnamefont
  {Kampen}},\ }\bibfield  {title} {\enquote {\bibinfo {title} {Quantum
  statistics of irreversible processes},}\ }\href {\doibase
  10.1016/S0031-8914(54)80074-7} {\bibfield  {journal} {\bibinfo  {journal}
  {Physica}\ }\textbf {\bibinfo {volume} {20}},\ \bibinfo {pages} {603--622}
  (\bibinfo {year} {1954})}\BibitemShut {NoStop}%
\bibitem [{\citenamefont {Dümcke}(1983)}]{DuemckeJMP1983}%
  \BibitemOpen
  \bibfield  {author} {\bibinfo {author} {\bibfnamefont {R.}~\bibnamefont
  {Dümcke}},\ }\bibfield  {title} {\enquote {\bibinfo {title} {Convergence of
  multitime correlation functions in the weak and singular coupling limits},}\
  }\href {\doibase 10.1063/1.525681} {\bibfield  {journal} {\bibinfo  {journal}
  {J. Math. Phys.}\ }\textbf {\bibinfo {volume} {24}},\ \bibinfo {pages} {311}
  (\bibinfo {year} {1983})}\BibitemShut {NoStop}%
\bibitem [{\citenamefont {Ford}\ and\ \citenamefont
  {O'Connell}(1996)}]{FordOConnellPRL1996}%
  \BibitemOpen
  \bibfield  {author} {\bibinfo {author} {\bibfnamefont {G.~W.}\ \bibnamefont
  {Ford}}\ and\ \bibinfo {author} {\bibfnamefont {R.~F.}\ \bibnamefont
  {O'Connell}},\ }\bibfield  {title} {\enquote {\bibinfo {title} {There is no
  quantum regression theorem},}\ }\href {\doibase 10.1103/PhysRevLett.77.798}
  {\bibfield  {journal} {\bibinfo  {journal} {Phys. Rev. Lett.}\ }\textbf
  {\bibinfo {volume} {77}},\ \bibinfo {pages} {798--801} (\bibinfo {year}
  {1996})}\BibitemShut {NoStop}%
\bibitem [{\citenamefont {Figueroa-Romero}\ \emph {et~al.}(2019)\citenamefont
  {Figueroa-Romero}, \citenamefont {Modi},\ and\ \citenamefont
  {Pollock}}]{FigueroaRomeroModiPollockQuantum2019}%
  \BibitemOpen
  \bibfield  {author} {\bibinfo {author} {\bibfnamefont {P.}~\bibnamefont
  {Figueroa-Romero}}, \bibinfo {author} {\bibfnamefont {K.}~\bibnamefont
  {Modi}}, \ and\ \bibinfo {author} {\bibfnamefont {F.~A.}\ \bibnamefont
  {Pollock}},\ }\bibfield  {title} {\enquote {\bibinfo {title} {Almost
  markovian processes from closed dynamics},}\ }\href {\doibase
  10.22331/q-2019-04-30-136} {\bibfield  {journal} {\bibinfo  {journal}
  {Quantum}\ }\textbf {\bibinfo {volume} {3}},\ \bibinfo {pages} {136}
  (\bibinfo {year} {2019})}\BibitemShut {NoStop}%
\bibitem [{\citenamefont {Figueroa-Romero}\ \emph {et~al.}(2021)\citenamefont
  {Figueroa-Romero}, \citenamefont {Pollock},\ and\ \citenamefont
  {Modi}}]{FigueroaRomeroPollockModiCP2021}%
  \BibitemOpen
  \bibfield  {author} {\bibinfo {author} {\bibfnamefont {P.}~\bibnamefont
  {Figueroa-Romero}}, \bibinfo {author} {\bibfnamefont {F.~A.}\ \bibnamefont
  {Pollock}}, \ and\ \bibinfo {author} {\bibfnamefont {K.}~\bibnamefont
  {Modi}},\ }\bibfield  {title} {\enquote {\bibinfo {title} {Markovianization
  with approximate unitary designs},}\ }\href {\doibase
  10.1038/s42005-021-00629-w} {\bibfield  {journal} {\bibinfo  {journal}
  {Commun. Phys.}\ }\textbf {\bibinfo {volume} {4}},\ \bibinfo {pages} {127}
  (\bibinfo {year} {2021})}\BibitemShut {NoStop}%
\bibitem [{\citenamefont {Strasberg}\ \emph {et~al.}(2023)\citenamefont
  {Strasberg}, \citenamefont {Winter}, \citenamefont {Gemmer},\ and\
  \citenamefont {Wang}}]{StrasbergEtAlPRA2023}%
  \BibitemOpen
  \bibfield  {author} {\bibinfo {author} {\bibfnamefont {P.}~\bibnamefont
  {Strasberg}}, \bibinfo {author} {\bibfnamefont {A.}~\bibnamefont {Winter}},
  \bibinfo {author} {\bibfnamefont {J.}~\bibnamefont {Gemmer}}, \ and\ \bibinfo
  {author} {\bibfnamefont {J.}~\bibnamefont {Wang}},\ }\bibfield  {title}
  {\enquote {\bibinfo {title} {Classicality, markovianity, and local detailed
  balance from pure-state dynamics},}\ }\href {\doibase
  10.1103/PhysRevA.108.012225} {\bibfield  {journal} {\bibinfo  {journal}
  {Phys. Rev. A}\ }\textbf {\bibinfo {volume} {108}},\ \bibinfo {pages}
  {012225} (\bibinfo {year} {2023})}\BibitemShut {NoStop}%
\bibitem [{\citenamefont {Strasberg}(2023)}]{StrasbergSP2023}%
  \BibitemOpen
  \bibfield  {author} {\bibinfo {author} {\bibfnamefont {P.}~\bibnamefont
  {Strasberg}},\ }\bibfield  {title} {\enquote {\bibinfo {title} {Classicality
  with(out) decoherence: Concepts, relation to markovianity, and a random
  matrix theory approach},}\ }\href {\doibase 10.21468/SciPostPhys.15.1.024}
  {\bibfield  {journal} {\bibinfo  {journal} {SciPost Phys.}\ }\textbf
  {\bibinfo {volume} {15}},\ \bibinfo {pages} {024} (\bibinfo {year}
  {2023})}\BibitemShut {NoStop}%
\bibitem [{\citenamefont {Strasberg}\ \emph {et~al.}(2024)\citenamefont
  {Strasberg}, \citenamefont {Reinhard},\ and\ \citenamefont
  {Schindler}}]{StrasbergReinhardSchindlerPRX2024}%
  \BibitemOpen
  \bibfield  {author} {\bibinfo {author} {\bibfnamefont {P.}~\bibnamefont
  {Strasberg}}, \bibinfo {author} {\bibfnamefont {T.~E.}\ \bibnamefont
  {Reinhard}}, \ and\ \bibinfo {author} {\bibfnamefont {J.}~\bibnamefont
  {Schindler}},\ }\bibfield  {title} {\enquote {\bibinfo {title} {First
  principles numerical demonstration of emergent decoherent histories},}\
  }\href {\doibase 10.1103/PhysRevX.14.041027} {\bibfield  {journal} {\bibinfo
  {journal} {Phys. Rev. X}\ }\textbf {\bibinfo {volume} {14}},\ \bibinfo
  {pages} {041027} (\bibinfo {year} {2024})}\BibitemShut {NoStop}%
\bibitem [{\citenamefont {Wang}\ and\ \citenamefont
  {Strasberg}(2025)}]{WangStrasbergArXiv2024}%
  \BibitemOpen
  \bibfield  {author} {\bibinfo {author} {\bibfnamefont {Jiaozi}\ \bibnamefont
  {Wang}}\ and\ \bibinfo {author} {\bibfnamefont {Philipp}\ \bibnamefont
  {Strasberg}},\ }\bibfield  {title} {\enquote {\bibinfo {title} {Decoherence
  of histories: Chaotic versus integrable systems},}\ }\href {\doibase
  10.1103/m8vq-l449} {\bibfield  {journal} {\bibinfo  {journal} {Phys. Rev.
  Lett.}\ }\textbf {\bibinfo {volume} {134}},\ \bibinfo {pages} {220401}
  (\bibinfo {year} {2025})}\BibitemShut {NoStop}%
\bibitem [{\citenamefont {Gaspard}\ and\ \citenamefont
  {Wang}(1993)}]{GaspardWangPR1993}%
  \BibitemOpen
  \bibfield  {author} {\bibinfo {author} {\bibfnamefont {P.}~\bibnamefont
  {Gaspard}}\ and\ \bibinfo {author} {\bibfnamefont {X.-J.}\ \bibnamefont
  {Wang}},\ }\bibfield  {title} {\enquote {\bibinfo {title} {Noise, chaos, and
  $(\epsilon,\tau)$-entropy per unit time},}\ }\href {\doibase
  10.1016/0370-1573(93)90012-3} {\bibfield  {journal} {\bibinfo  {journal}
  {Phys. Rep.}\ }\textbf {\bibinfo {volume} {235}},\ \bibinfo {pages}
  {291--343} (\bibinfo {year} {1993})}\BibitemShut {NoStop}%
\bibitem [{\citenamefont {Debnath}\ \emph {et~al.}(2020)\citenamefont
  {Debnath}, \citenamefont {Dold}, \citenamefont {Morton},\ and\ \citenamefont
  {Mølmer}}]{DebnathEtAlPRL2020}%
  \BibitemOpen
  \bibfield  {author} {\bibinfo {author} {\bibfnamefont {K.}~\bibnamefont
  {Debnath}}, \bibinfo {author} {\bibfnamefont {G.}~\bibnamefont {Dold}},
  \bibinfo {author} {\bibfnamefont {J.~J.~L.}\ \bibnamefont {Morton}}, \ and\
  \bibinfo {author} {\bibfnamefont {K.}~\bibnamefont {Mølmer}},\ }\bibfield
  {title} {\enquote {\bibinfo {title} {Self-stimulated pulse echo trains from
  inhomogeneously broadened spin ensembles},}\ }\href {\doibase
  10.1103/PhysRevLett.125.137702} {\bibfield  {journal} {\bibinfo  {journal}
  {Phys. Rev. Lett.}\ }\textbf {\bibinfo {volume} {125}},\ \bibinfo {pages}
  {137702} (\bibinfo {year} {2020})}\BibitemShut {NoStop}%
\bibitem [{\citenamefont {Weichselbaumer}\ \emph {et~al.}(2020)\citenamefont
  {Weichselbaumer}, \citenamefont {Zens}, \citenamefont {Zollitsch},
  \citenamefont {Brandt}, \citenamefont {Rotter}, \citenamefont {Gross},\ and\
  \citenamefont {Huebl}}]{WeichselbaumerEtAlPRL2020}%
  \BibitemOpen
  \bibfield  {author} {\bibinfo {author} {\bibfnamefont {S.}~\bibnamefont
  {Weichselbaumer}}, \bibinfo {author} {\bibfnamefont {M.}~\bibnamefont
  {Zens}}, \bibinfo {author} {\bibfnamefont {C.~W.}\ \bibnamefont {Zollitsch}},
  \bibinfo {author} {\bibfnamefont {M.~S.}\ \bibnamefont {Brandt}}, \bibinfo
  {author} {\bibfnamefont {S.}~\bibnamefont {Rotter}}, \bibinfo {author}
  {\bibfnamefont {R.}~\bibnamefont {Gross}}, \ and\ \bibinfo {author}
  {\bibfnamefont {H.}~\bibnamefont {Huebl}},\ }\bibfield  {title} {\enquote
  {\bibinfo {title} {Echo trains in pulsed electron spin resonance of a
  strongly coupled spin ensemble},}\ }\href {\doibase
  10.1103/PhysRevLett.125.137701} {\bibfield  {journal} {\bibinfo  {journal}
  {Phys. Rev. Lett.}\ }\textbf {\bibinfo {volume} {125}},\ \bibinfo {pages}
  {137701} (\bibinfo {year} {2020})}\BibitemShut {NoStop}%
\end{thebibliography}%

\end{document}